\documentclass[lettersize,journal]{IEEEtran}
\usepackage{amsmath,amsfonts}
\usepackage{algorithmic}
\usepackage{array}
\usepackage[caption=false,font=normalsize,labelfont=sf,textfont=sf]{subfig}
\usepackage{color,soul}
\usepackage{textcomp}
\usepackage{cite}
\usepackage{stfloats}
\usepackage{url}
\usepackage{enumitem}
\usepackage{multirow}
\usepackage{verbatim}
\usepackage{graphicx}
\usepackage{hyperref}
 
\def\BibTeX{{\rm B\kern-.05em{\sc i\kern-.025em b}\kern-.08em
    T\kern-.1667em\lower.7ex\hbox{E}\kern-.125emX}}
\usepackage{balance}
\begin{document}
\title{Joint Communications and Sensing for 6G Satellite Networks: Use Cases and Challenges}
\author{Chandan Kumar Sheemar, Prabhu Thiruvasagam, Wali Ullah Khan, Sourabh Solanki,\\ George C. Alexandropoulos, Jorge Querol, 
Jan Plachy, Oliver Holschke, and Symeon Chatzinotas
\thanks{The auhtors Chandan Kumar Sheemar, Prabhu Thiruvasagam, Wali Ullah Khan, Jorge Querol and Symeon Chatzinotas are with the SnT department at the University of Luxembourg (email:\{chandankumar.sheemar, prabhu.thiruvasagam,jorge.querol,symeon.chatzinotas\}@uni.lu); Sourabh Solanki is
with the Department of ECE, National Institute of Technology Warangal,
TS, 506004, India, E-mail: (ssolanki@nitw.ac.in); G. C. Alexandropoulos is with
the Department of Informatics and Telecommunications, National and
Kapodistrian University of Athens, 16122 Athens, Greece, and with the
Department of Electrical and Computer Engineering, University of Illinois
Chicago, IL 60601, USA {email: alexandg@di.uoa.gr}; Jan Plachy and Oliver Holschke are with Deutsche Telekom, Germany, email\{jan.plachy@telekom.de,oliver.holschke@telekom.de\} }
}
\markboth{}%
{How to Use the IEEEtran \LaTeX \ Templates}

\maketitle

 \begin{abstract}
Satellite networks (SN) have long provided two fundamental services: global communications and Earth-oriented sensing, supporting applications from connectivity and navigation to disaster management and environmental monitoring. Yet, the accelerating demand for data and the emergence of new applications render the independent evolution of communication and sensing payloads increasingly unsustainable. Joint communications and sensing (JCAS) has emerged as a transformative paradigm, integrating both functions within a unified payload to enhance spectral efficiency, reduce operational costs, and minimize hardware redundancy. Beyond efficiency, this integration creates opportunities for novel services that are infeasible under separate payload designs. This paper motivates the role of JCAS in shaping the sixth-generation (6G) of satellite networks, explores a representative use case to assess its feasibility, and discusses key challenges that must be addressed to unlock its full potential. By highlighting these opportunities inherent to the space environment,
we aim to stimulate the development of JCAS as a cornerstone
technology for the next-generation space era.
\end{abstract}


\section{Introduction}
\IEEEPARstart{S}{atellite Networks} (SN) represent a paradigm-shifting advancement in communication technology, aiming to transcend the limitations of traditional terrestrial infrastructures \cite{10396846}. In stark contrast to conventional networks that rely on ground-based infrastructures for their operations and providing services, SN employs a sophisticated amalgamation of satellites at different altitude encompassing Low Earth Orbit (LEO), Medium Earth Orbit (MEO), and Geostationary Equatorial Orbit (GEO) to establish pervasive coverage and uninterrupted connectivity on a global scale \cite{9210567}. SN have the capacity to revolutionize connectivity in remote and underserved regions, facilitate swift communication restoration in disaster-stricken areas, and even extend their influence to the vast expanses of the oceans and outer space \cite{giordani2020non}. Currently, SN provide two primary services: i) communications and ii) sensing. In particular, $> 8000$ satellites are dedicated to communications and data services and $>1000$ satellites are dedicated to support sensing applications \cite{nanoavionics2024many}.

 In parallel to the advancement of SN, the emergence of joint communications and sensing (JCAS) technology represents a momentous leap in wireless systems \cite{chepuri2023integrated,sheemar2023full}. They aim to seamlessly amalgamate communication and sensing functionalities, allowing a single system to transmit data while simultaneously collecting critical environmental information. This integration not only optimizes spectral and energy resources but also enhances the efficiency and versatility of wireless networks by reducing the need for separate systems dedicated solely to either communication or sensing \cite{zhang2021overview,sheemar2023full_magazine}. Beyond efficiency, JCAS enables novel services that rely on real-time situational awareness, from intelligent transportation to space traffic management. It also lays the groundwork for future infrastructures where connectivity and sensing are no longer parallel functions but inseparable components of a unified system \cite{sheemar2025holographic,9737357}.


Integrating JCAS technology into SN marks a transformative step toward redefining the next generation of global wireless networks. This advancement promises a range of powerful capabilities, from ultra-fast data transmission and broader coverage in remote and underserved areas to precise sensing functions at the global scale. Recent works on this topic are available in \cite{leyva2024integrated,naeem2022novel,yin2024integrated,you2022beam}. In \cite{leyva2024integrated}, the use case of continent-wide JCAS with multiple LEO satellites is investigated. The authors propose a JCAS framework to precisely estimate attenuation in communication links caused by precipitation, aiming to identify optimal serving satellites and allocate resources efficiently for downlink communication with ground-based users. In \cite{naeem2022novel}, a novel framework for JCAS is proposed to develop a synergy between the SN and the terrestrial networks (TNs). The proposed design addresses the limitations of conventional time division duplexing (TDD) mode by enabling communication during the radar pulse's waiting period, thereby minimizing latency between sensing and communication.
In \cite{yin2024integrated}, the authors explore the rate-splitting approach as a potential solution for JCAS in Low LEO SN. In \cite{you2022beam}, a novel hybrid beamforming approach for massive MIMO LEO satellite systems is presented. 

\begin{figure*}
    \centering
    \includegraphics[width=\linewidth]{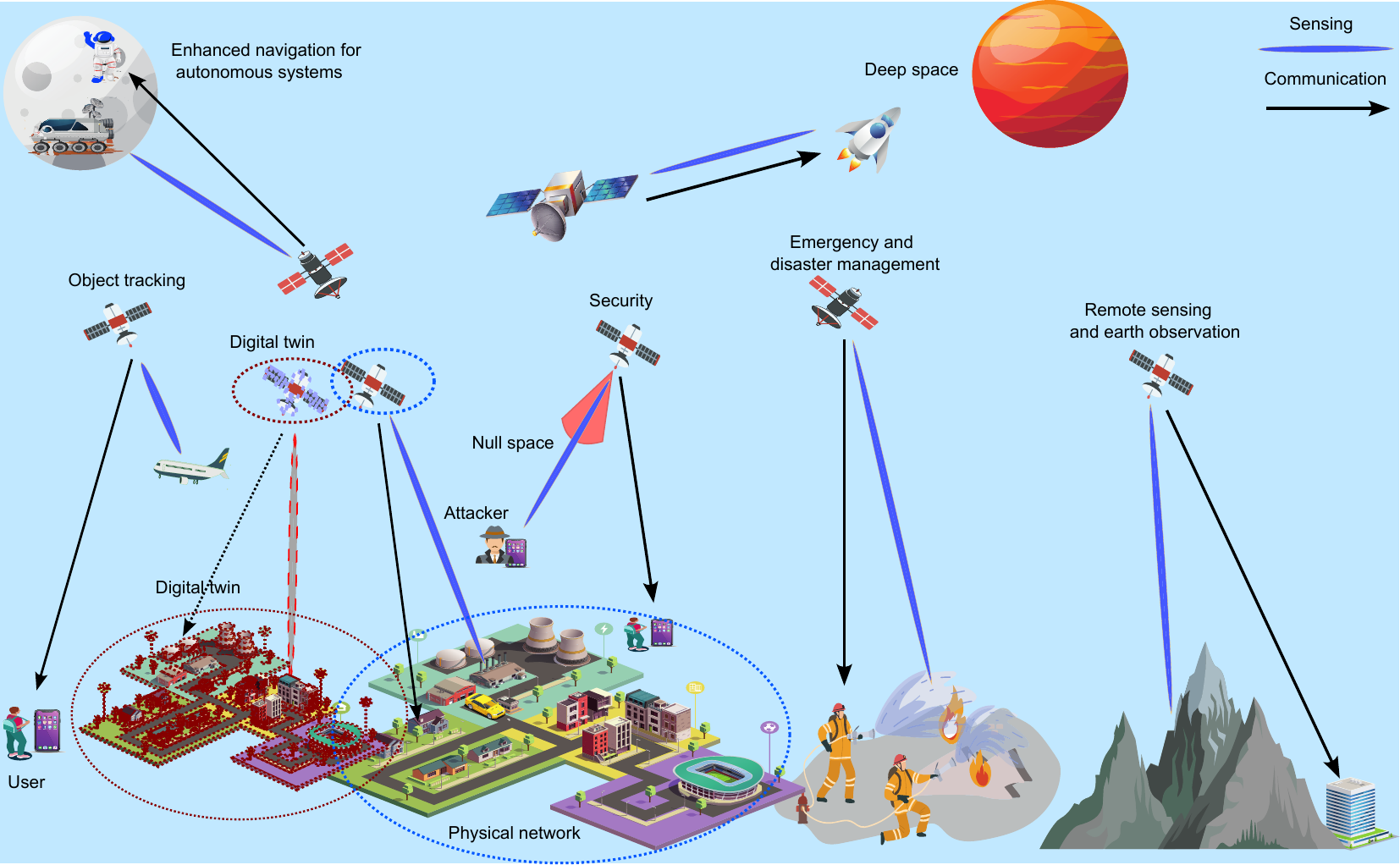}
    \caption{Novel sensing use cases enabled jointly with communications for next-generation SN.}
    \label{fig:HC}
\end{figure*}

Although JCAS offers immense potential to extend global coverage and unlock new services in satellite networks (SN), research on its integration into space systems is still in its early stages. Existing studies provide promising insights but remain limited in scope, leaving many practical and theoretical challenges unresolved. This article aims to stimulate further exploration by introducing innovative use cases for JCAS in SN, shown in Figure \ref{fig:HC}, which can be enabled with an interplay between communications and sensing.
Many of these use cases would be infeasible with independent payload designs, or at best achievable with limited performance and without real-time capability, underscoring the need for integrated solutions. To demonstrate the feasibility of JCAS in space, one representative use case is studied in detail. Finally, the paper discusses key challenges inherent to JCAS operation in space and outlines promising directions for future research.

\emph{Paper Organization:} In the following, Section \ref{sezione2} presents the preliminaries and overview of SN and motivates the potential of JCAS. Section~\ref{sezione3} presents the new use cases, and Section~\ref{sezione4} presents the case study. Finally, Sections~\ref{sezione5} and \ref{sezione6} present the open challenges and conclusions.

\section{Preliminaries and Motivation} \label{sezione2}

This section presents an overview of the key principles and driving factors behind the development of next-generation SN, including architecture, payloads, and the motivation for the convergence of communications and sensing services into a unified payload.

\subsection{SN Architecture}
Hereafter, we provide an overview of the typical SN architecture consisting of three main segments.
\subsubsection{Space Segment} 
In SN, this segment refers to the collection of satellites and associated equipment operating in orbit as part of the overall system \cite{maral2020satellite}. It consists of several key components: a) \emph{Satellites or Spacecraft}, which serve as the core units in GEO, MEO, or LEO and carry instruments for communication, sensing, or scientific missions; b) \emph{Payload}, which performs the primary mission functions. For communications, payloads often employ either transparent repeaters (bent-pipe), which simply amplify and forward signals at a different frequency, or regenerative repeaters, which demodulate the uplink signal to baseband and re-modulate it for downlink transmission. For sensing, payloads gather imagery, environmental parameters, or scientific measurements. c) \emph{Inter-Satellite Links (ISLs)}, which allow satellites in a constellation to exchange data directly, improving coverage, redundancy, and latency performance; and d) \emph{On-board Processing Systems}, which provide capabilities such as filtering, compression, and traffic management, reducing the volume of data transmitted to Earth and enhancing overall efficiency.

\subsubsection{Ground Segment} 
The segment is the Earth-based component of a SN, encompassing all infrastructure required to operate, monitor, and utilize satellites in orbit. It consists of: a) \emph{Ground Stations}, which use large antennas and communication systems to transmit commands to satellites and receive downlinked information, including communication signals and sensing data; b) \emph{Control Centers}, which oversee satellite operations through telemetry, tracking, and command (TTC). These centers monitor satellite health, orbital parameters, and system performance, and issue commands for orbit adjustments, attitude control, and payload operations; c) \emph{Data Processing Centers}, where received information is decoded, analyzed, and transformed into usable products such as Earth observation imagery, environmental models, or communication services; and d) \emph{Network Infrastructure}, which links ground stations, control centers, and data facilities through fiber, dedicated backhaul, or the internet to ensure secure and efficient distribution of satellite data to end users and applications.



\subsubsection{User Segment} The user segment is the part of a SN that includes all devices, equipment, and technology used by end-users to access services provided by the SN. Essentially, it’s the interface between the satellite system and the end-users, enabling them to receive data, communicate, or utilize satellite-based services. 

\begin{table}
\centering
\caption{Communication satellites frequency bands.} 
\begin{tabular}{ |p{1.3cm}|p{1.4cm}|p{4.5cm}|}
\hline
Freq. Band & Range(GHz)& Traditional Applications \\
\hline\hline
L-Band  & 1.518-1.675    & Civil mobile communications, global positioning
systems,
and weather radar systems \\\hline
S-Band &   1.97-2.69  & Satellite TV, mobile broadband services, radio,
and in-flight connectivity  \\\hline
C-Band  &3.4-7.025 & Device services, satellite TV, unprocessed satellite
feeds  \\\hline
X-Band & 7.25-8.44 & Military operations, pulse radar system, weather
monitoring, air traffic control \\\hline
Ku-Band &  10.7-14.5 & Fixed satellite television data services \\\hline
Ka-Band & 17.3-30 & Two-way broadband services, fixed satellite data
services \\\hline
Q-V-Band  & 37.5-51.4  & High speed broadband services, in-flight connectivity \\
\hline 
\end{tabular} 
\label{sat_app}
\end{table}

\begin{table*}[]
\centering
\caption{Frequency bands allocated for spaceborne radars}
\label{Table:Frequency Band Radar}
\begin{tabular}{|c|l|lllll|}
\hline
\multirow{2}{*}{\begin{tabular}[c]{@{}c@{}}Satellite Frequency \\ Bands\end{tabular}} & \multicolumn{1}{c|}{\multirow{2}{*}{\begin{tabular}[c]{@{}c@{}}Allocated Frequency\\ for Active Sensing\end{tabular}}} & \multicolumn{5}{c|}{Assigned Bandwidth for Sensors}                                                                                                                                                                                                                               \\ \cline{3-7} 
                                                                                      & \multicolumn{1}{c|}{}                                                                                                  & \multicolumn{1}{c|}{Scatterometer} & \multicolumn{1}{c|}{Altimeter} & \multicolumn{1}{c|}{SAR}        & \multicolumn{1}{c|}{\begin{tabular}[c]{@{}c@{}}Precipitation \\ Radar\end{tabular}} & \multicolumn{1}{c|}{\begin{tabular}[c]{@{}c@{}}Cloud Profile\\  Radar\end{tabular}} \\ \hline
P (0.3-1 GHz)                                                                         & 432-438 MHz                                                                                                            & \multicolumn{1}{l|}{}              & \multicolumn{1}{l|}{}          & \multicolumn{1}{l|}{6 MHz}      & \multicolumn{1}{l|}{}                                                               &                                                                                     \\ \hline
L (1-2 GHz)                                                                           & 1215-1300 MHz                                                                                                          & \multicolumn{1}{l|}{5-500 kHz}     & \multicolumn{1}{l|}{}          & \multicolumn{1}{l|}{20-85 MHz}  & \multicolumn{1}{l|}{}                                                               &                                                                                     \\ \hline
S (2-4 GHz)                                                                           & 3100-3300 MHz                                                                                                          & \multicolumn{1}{l|}{}              & \multicolumn{1}{l|}{200 MHz}   & \multicolumn{1}{l|}{20-200 MHz} & \multicolumn{1}{l|}{}                                                               &                                                                                     \\ \hline
C (4-8 GHz)                                                                           & 5250-5570 MHz                                                                                                          & \multicolumn{1}{l|}{5-500 kHz}     & \multicolumn{1}{l|}{320 MHz}   & \multicolumn{1}{l|}{20-320 MHz} & \multicolumn{1}{l|}{}                                                               &                                                                                     \\ \hline
\multirow{2}{*}{X (8-12 GHz)}                                                         & 8550-8650 MHz                                                                                                          & \multicolumn{1}{l|}{5-500 kHz}     & \multicolumn{1}{l|}{100 MHz}   & \multicolumn{1}{l|}{20-100 MHz} & \multicolumn{1}{l|}{}                                                               &                                                                                     \\ \cline{2-7} 
                                                                                      & 9300-9900 MHz                                                                                                          & \multicolumn{1}{l|}{5-500 kHz}     & \multicolumn{1}{l|}{300 MHz}   & \multicolumn{1}{l|}{20-600 MHz} & \multicolumn{1}{l|}{}                                                               &                                                                                     \\ \hline
\multirow{2}{*}{Ku (12-18 GHz)}                                                       & 13.25-13.75 GHz                                                                                                        & \multicolumn{1}{l|}{5-500 kHz}     & \multicolumn{1}{l|}{500 MHz}   & \multicolumn{1}{l|}{}           & \multicolumn{1}{l|}{0.6-14 MHz}                                                     &                                                                                     \\ \cline{2-7} 
                                                                                      & 17.2-17.3 GHz                                                                                                          & \multicolumn{1}{l|}{5-500 kHz}     & \multicolumn{1}{l|}{}          & \multicolumn{1}{l|}{}           & \multicolumn{1}{l|}{0.6-14 MHz}                                                     &                                                                                     \\ \hline
K (18-27 GHz)                                                                         & 24.05-24.25 GHz                                                                                                        & \multicolumn{1}{l|}{}              & \multicolumn{1}{l|}{}          & \multicolumn{1}{l|}{}           & \multicolumn{1}{l|}{0.6-14 MHz}                                                     &                                                                                     \\ \hline
Ka (27-40 GHz)                                                                        & 35.5-36 GHz                                                                                                            & \multicolumn{1}{l|}{5-500 kHz}     & \multicolumn{1}{l|}{500 MHz}   & \multicolumn{1}{l|}{}           & \multicolumn{1}{l|}{0.6-14 MHz}                                                     &                                                                                     \\ \hline
\multirow{2}{*}{W (75-110 GHz)}                                                       & 78-79 GHz                                                                                                              & \multicolumn{1}{l|}{}              & \multicolumn{1}{l|}{}          & \multicolumn{1}{l|}{}           & \multicolumn{1}{l|}{}                                                               & 0.3-10 MHz                                                                          \\ \cline{2-7} 
                                                                                      & 94-94.1 GHz                                                                                                            & \multicolumn{1}{l|}{}              & \multicolumn{1}{l|}{}          & \multicolumn{1}{l|}{}           & \multicolumn{1}{l|}{}                                                               & 0.3-10 MHz                                                                          \\ \hline
\multicolumn{1}{|l|}{\multirow{2}{*}{G (110-300 GHz)}}                                & 133.5-134 GHz                                                                                                          & \multicolumn{1}{l|}{}              & \multicolumn{1}{l|}{}          & \multicolumn{1}{l|}{}           & \multicolumn{1}{l|}{}                                                               & 0.3-10 MHz                                                                          \\ \cline{2-7} 
\multicolumn{1}{|l|}{}                                                                & 237.9-238 GHz                                                                                                          & \multicolumn{1}{l|}{}              & \multicolumn{1}{l|}{}          & \multicolumn{1}{l|}{}           & \multicolumn{1}{l|}{}                                                               & 0.3-10 MHz                                                                          \\ \hline
\end{tabular}
\end{table*}
\subsection{Communications Payloads}
Communication satellites serve as essential infrastructure in modern telecommunications, with their payloads broadly comprising two main subsystems: the antenna subsystem and the on-board digital processor (OBDP). The antenna subsystem is increasingly based on advanced direct-radiating phased array architectures, enabling flexible and high-capacity connectivity. These arrays support analog, digital, or hybrid beamforming techniques, with hybrid beamforming emerging as a highly effective solution to address challenges such as limited packaging space, power efficiency, and constraints in digital processing capabilities. Complementing the antenna system, the OBDP functions as the central digital signal processing unit onboard the satellite. It is responsible for executing critical signal processing tasks—such as encoding, decoding, modulation, and error correction—to ensure data integrity and transmission reliability. In addition, the OBDP oversees vital payload management functions, including dynamic frequency assignment, power allocation, and system resource optimization, all of which contribute to enhancing the overall performance and adaptability of satellite communication systems. Table~\ref{sat_app} highlights the frequency bands allocated for the communication satellites.

\subsection{Sensing Payloads}
Spaceborne remote sensing provides an unparalleled vantage point for observing the Earth, offering critical insights into the planet’s atmosphere, land, and oceans. Among the most advanced tools in this domain are spaceborne radars—active sensors that transmit electromagnetic waves from orbit and measure the backscattered signals from the Earth's surface. These measurements enable a wide range of applications including environmental monitoring, weather forecasting, pollution detection, and climate research. There are five principal types of spaceborne radar systems: scatterometers, altimeters, synthetic aperture radar (SAR), precipitation radars, and cloud profiling radars. Each of these sensors is tailored to measure distinct physical properties and, as a result, operates over specific frequency bands and bandwidths to meet their resolution requirements, as summarized in Table \ref{Table:Frequency Band Radar}.

The radar payload supporting these systems is typically divided into two primary subsystems: the Radio Frequency Unit (RFU) and the Digital Processing Unit (DPU). The RFU is responsible for the RF signal chain, including key components such as low-pass filters (LPFs), power amplifiers (PAs), low-noise amplifiers (LNAs), mixers, and local oscillators (LOs) for frequency conversion. It also incorporates RF chains, beamforming networks to shape and direct the radar beams, and duplexers to enable bidirectional transmission through a single antenna. Complementing the RFU, the DPU is in charge of generating radar waveforms using digital generators, converting digital signals to analog via DACs for transmission, and receiving and digitizing echo signals via ADCs. The DPU also runs detection and estimation algorithms, formats the processed data into interpretable images, and includes a control unit to manage subsystem operations. Echo signals received by the antenna are processed through the LNA and frequency down-converted, filtered, and amplified before digital processing. The resulting Earth observation (EO) data is then transmitted to ground stations for further analysis and delivery of value-added services.

\subsection{Motivation for JCAS in SN}
 The integration of JCAS into SN marks a transformative step toward multi-mission space platforms that can operate under the strict size, weight, power, and thermal (SWaP-T) constraints inherent to satellites. By consolidating communication and sensing into a common hardware and spectrum framework, JCAS payloads can achieve extremely high spectral efficiency through the use of joint waveforms and shared frequency bands, enabling simultaneous transmission and sensing without additional allocations. This integration improves system capacity and situational awareness at the global scale, while also streamlining payload architectures by reducing the need for duplicated antennas, transceivers, and baseband processors. As a result, JCAS enhances both throughput and sensing fidelity, while directly addressing spectrum scarcity and resource limitations in the space domain.

Beyond efficiency, JCAS provides tangible economic and sustainability benefits for satellite operators. Payload consolidation reduces the overall mass and volume of spacecraft, thereby lowering satellite manufacturing costs and launch expenses. The reduction in SWaP-T requirements not only improves energy efficiency but also allows satellites to achieve longer operational lifetimes or to allocate resources to additional mission capabilities. At the constellation level, fewer satellites are needed to provide equivalent coverage and sensing refresh rates, which decreases both capital expenditures (CAPEX) and operational expenditures (OPEX) across the satellite lifecycle. Moreover, integrating functions into fewer spacecraft reduces the risk of collision and helps mitigate the long-term challenge of space debris. Finally, shared hardware platforms lower bill-of-materials and qualification costs, while sensing-enhanced channel awareness improves adaptive resource allocation and beam management. Collectively, these advantages make JCAS a cornerstone for the development of cost-effective, energy-efficient, and sustainable next-generation satellite networks.

\section{JCAS for 6G SN: Potential Use Cases} \label{sezione3}
In the following sections, we explore the capabilities of JCAS-powered SN to enable novel use cases.   

\subsection{Joint Communications and Remote sensing and/or Earth observation (EO)}

 Traditionally, remote sensing or EO satellites collect data and then downlink it in batches, often introducing latency between observation and actionable insight. Furthermore, sensing tasks operate on pre-allocated frequency bands, limiting flexibility in adapting to dynamic environmental conditions. By embedding JCAS capabilities into SN, these limitations can be addressed directly. Real-time integration of communication and sensing allows satellites to transmit critical Earth observation data as it is acquired, significantly reducing decision-making latency in time-sensitive scenarios such as wildfire progression, flood monitoring, or storm tracking. At the same time, communication signals are jointly designed as sensing sources, providing required sensing operations over different geographical areas. This dual role enhances coverage across terrestrial, marine, and atmospheric environments, enabling continuous situational awareness with finer spatial and temporal resolution. Moreover, JCAS can facilitate adaptive spectrum use, where sensing observations inform communication link adaptation (e.g., precipitation-induced attenuation), thereby improving both EO fidelity and communication reliability.

\subsection{Joint Communication and Navigation for Autonomous Systems} 

 Future space exploration will increasingly depend on autonomous systems including rovers, landers, and spacecraft, that must navigate and operate with minimal human intervention in unfamiliar extraterrestrial environments. JCAS technology has the potential to become the backbone of such extraterrestrial autonomy. Communication signals transmitted between satellites and surface platforms can be jointly designed as navigation and sensing tools, enabling rovers to simultaneously receive data and perceive their environment. In practice, this means a rover could not only download mission commands but also detect nearby obstacles, identify other rovers or landers, and update its position with high precision. By providing a shared situational awareness layer, JCAS can enable coordinated traffic management, swarm-based exploration, and collaborative construction activities on planetary surfaces. Looking ahead, such capabilities could form the foundation of \emph{off-world intelligent transportation systems}, where fleets of autonomous explorers navigate safely and cooperatively, accelerating the establishment of sustainable human and robotic presence on the Moon, Mars, and beyond.


\subsection{Joint Communication and Security} 
 
Conventional systems treat sensing and communication separately, leaving communication channels vulnerable to falsified signals and limiting the ability to independently verify transmitted information. By merging the two functions, JCAS-SN establishes a dual layer of defense where sensing-derived insights can directly strengthen the security of communication links. In practice, this means that signals exchanged through systems such as the Automatic Identification System (AIS), the VHF Data Exchange System (VDES), and Automatic Dependent Surveillance-Broadcast (ADS-B) can be continuously cross-checked against independent sensing measurements. If, for example, a vessel transmits a falsified AIS position, JCAS can verify its true trajectory through sensing, immediately exposing the spoofing attempt. At the same time, the communication channel itself can benefit from this integration: sensing information can be used to adaptively authenticate links, localize malicious emitters, and trigger protective measures such as beam nulling or targeted jamming to secure the spectrum. Beyond preventing unauthorized access, this synergy ensures both the fidelity of navigation-related data and the robustness of communication services.

\subsection{Joint Communication and Object Detection}

JCAS-enabled SN have the potential to deliver a step-change in object detection by transforming connectivity into a precision sensing asset. Multistatic geometries formed by distributed LEO, MEO, and GEO nodes can be leveraged to enable continuous illumination and observation of targets, where the same waveforms that sustain communication links also provide bi-/multistatic range, Doppler, and angle estimates for high-accuracy detection and kinematic tracking. For cooperative targets, not only uplink and downlink pilots but also jointly designed data signaling can be exploited to achieve centimeter-to-decimeter localization through time-difference-of-arrival (TDOA), frequency-difference-of-arrival (FDOA), and phase-tracking techniques. For non-cooperative targets such as ships beyond coastal coverage, aircraft outside radar networks, or orbital debris, reflections of communication signals can be utilized with coherent integration, clutter suppression over ocean or ice, and elevation diversity to extend detection range and reliability. On-board edge processing can further fuse returns across inter-satellite links, enabling continuous tracking, ephemeris refinement, and advanced classification through micro-Doppler or polarimetric analysis. In turn, the resulting sensing outputs can drive adaptive communication functions: beams can hop or steer to follow high-mobility targets, safety-critical traffic can be prioritized through quality-of-service (QoS) mechanisms, and alerts can be disseminated to other satellites or ground catalogs for conjunction assessment and planetary-defense screening.


\subsection{Joint Communication and Digital Twin}
JCAS-enabled SN can play a central role in the development of digital twins (DTs) by unifying dual services into a unified framework. On the sensing side, JCAS can enable satellites to collect multi-modal data, including microwave returns, reflected communication signals, and imagery, capturing not only surface and environmental conditions on Earth but also the terrain of the Moon or other celestial bodies. On the communication side, these same links can ensure the continuous transfer of sensed data to ground stations or directly between satellites, enabling real-time synchronization of DTs without dependence on separate data-acquisition or channel-estimation processes. By combining these dual functions, JCAS-based SN can construct and update highly accurate digital replicas of physical environments, supporting predictive modeling, simulation, and remote monitoring at a planetary scale. Such DTs can be used for climate forecasting, disaster preparedness, orbital environment management, and mission planning in extraterrestrial domains. Moreover, the tight integration of communication and sensing can enable predictive analytics to anticipate environmental changes while simultaneously adjusting network resource allocation, ensuring resilient operation under dynamic and challenging conditions.

\subsection{Joint Communication and Disaster Management}
JCAS-enabled SN can uniquely couple hazard sensing with time-critical communications to support end-to-end disaster management when terrestrial infrastructure is impaired or absent. On the sensing side, communication waveforms and microwave returns can be exploited to detect and characterize events such as extreme rainfall, flood extent, storm tracks, wildfire fronts, landslides, and infrastructure damage—using attenuation, Doppler, and reflection signatures, as well as imaging and change detection. Device-based measurements (e.g., from handsets, AIS/VDES beacons, or ADS-B) can complement device-free sensing to localize people and vessels in affected areas, which can be further enforced with satellite-borne positioning for georeferenced situational awareness for first responders. On the communications side, the jointly designed waveforms can be used to deliver authenticated alerts and guidance via real-time broadcast/multicast/geocast to the impacted footprint, prioritizing emergency traffic with QoS and establishing temporary coverage (direct-to-device or via aerial relays) where ground networks are down. Sensing outputs can then be used to also dynamically steer beams in real-time and allocate spectrum toward evolving hotspots; in return, crowd-sourced link metrics and terminal reports can enhance hazard maps and search areas. 


\subsection{Deep Space}
In deep-space missions spanning cislunar space to planetary, small-body, and interstellar precursors, JCAS-enabled SN can unify long-range communications and scientific/proximity sensing under the same signaling fabric, addressing constraints intrinsic to vast distances, sparse infrastructure, and extreme light-time delay. On the sensing side, communication waveforms themselves can be repurposed for bi-/multistatic radar sounding of planetary surfaces, regolith, and ring systems. Radiometric navigation and differential techniques can provide precise ephemeris and relative-state estimation during cruise, approach, and proximity operations, while onboard processing can fuses crosslink returns to maintain continuous tracks of the regions in deep space. On the communications side, sensing outputs can directly inform link adaptation under ultra-low SNR and high Doppler dynamics: beams can be steered and scheduled around occultations; coding/modulation and power are adapted to predicted scintillation; and delay-tolerant networking (DTN) policies can be triggered by event detections for priority, event-driven downlink. Crosslinks among relay spacecraft can create distributed apertures that both extend receive sensitivity and enable multistatic geometries for simultaneous science and telemetry. This tight sensing–communications loop can yield resilient, low-latency deep-space situational awareness despite long round-trip times, enabling hazard detection, autonomous trajectory updates, and time-critical science capture that would be infeasible with disjoint systems.

 
\begin{figure}
    \centering
\includegraphics[width=0.5\textwidth]{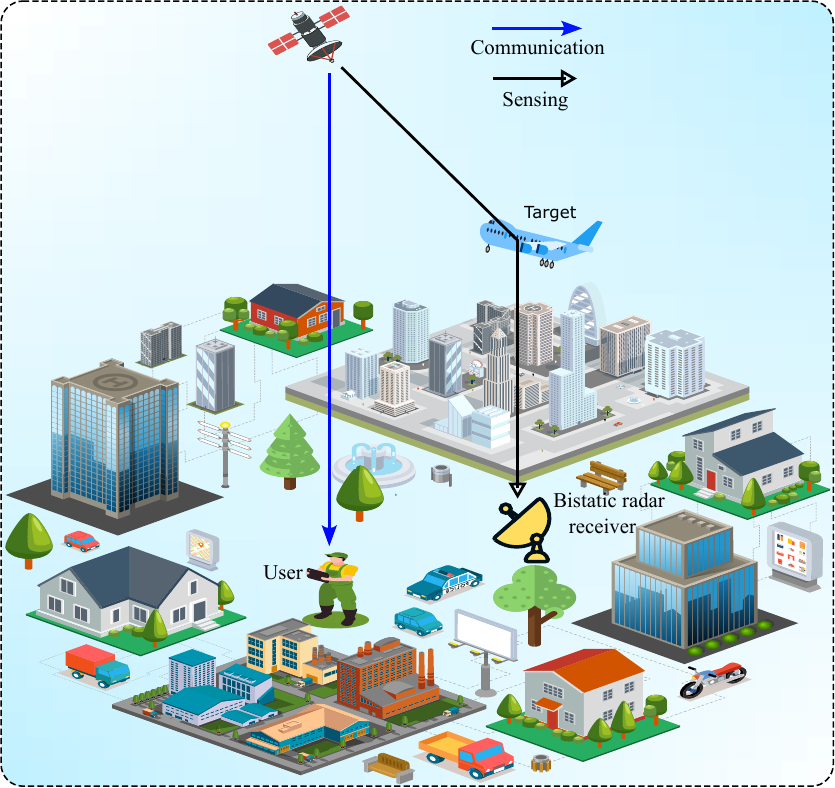}
    \caption{Selected use case for JCAS SN.}
    \label{JCAS_Chandan}
\end{figure} 
\section{A Next-Generation JCAS SN: A Case Study} \label{sezione4}
In this section, we aim to study the feasibility of the joint communications and object detection use case. We simulate a bistatic JCAS LEO satellite system operating at $4.2$ GHz (C-band), where the satellite transmits signals while a ground-based receiver captures radar reflections and receives communications data. The system features a $500$ km satellite-to-user distance for communications and a $490$ km satellite-to-aircraft distance combined with a $10$ km aircraft-to-ground-receiver distance for bistatic radar detection. The aircraft target has a radar cross-section (RCS) of $100 m^2$, and the system utilizes a $100$ MHz bandwidth OFDM waveform with $1024$ subcarriers ($800$ data subcarriers, and $224$ for sensing) and a $72$-sample cyclic prefix, employing QPSK modulation for robust communications. The transmit and receive antenna gains are assumed to be $22.81$ dBi and $32.85$ dBi, respectively. Radar processing uses a $300$ ms integration time, and the system operates across a $1-9$ dBW transmit power range, with thermal noise at $300$ K. 
The user is assumed to be located at the $10^\circ$ elevation angle, and the target at the $30^\circ$ elevation angles. We assume the Doppler effect to be precompensated.

 \begin{figure}
     \centering
     \includegraphics[width=0.8\linewidth]{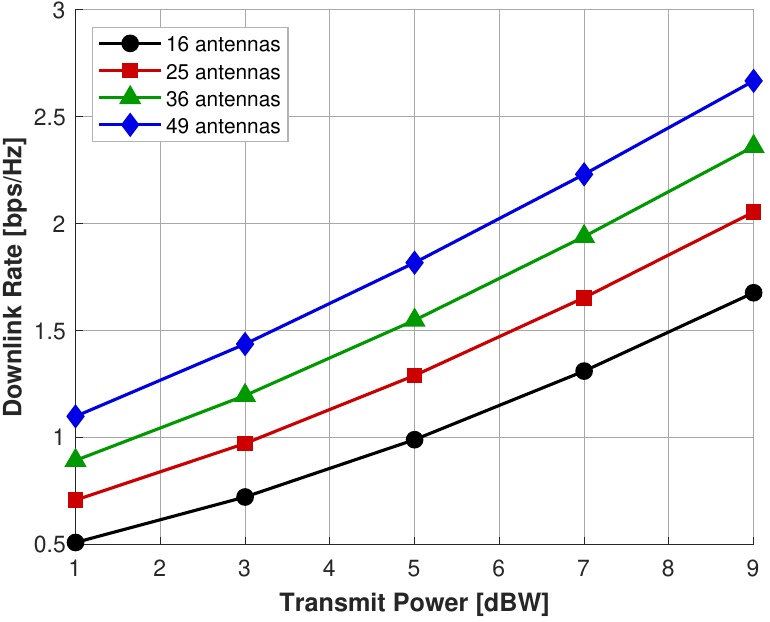}
     \caption{Communications performance as a function of transmit power.}
     \label{fig4}
 \end{figure}

  \begin{figure}
     \centering
     \includegraphics[width=0.8\linewidth]{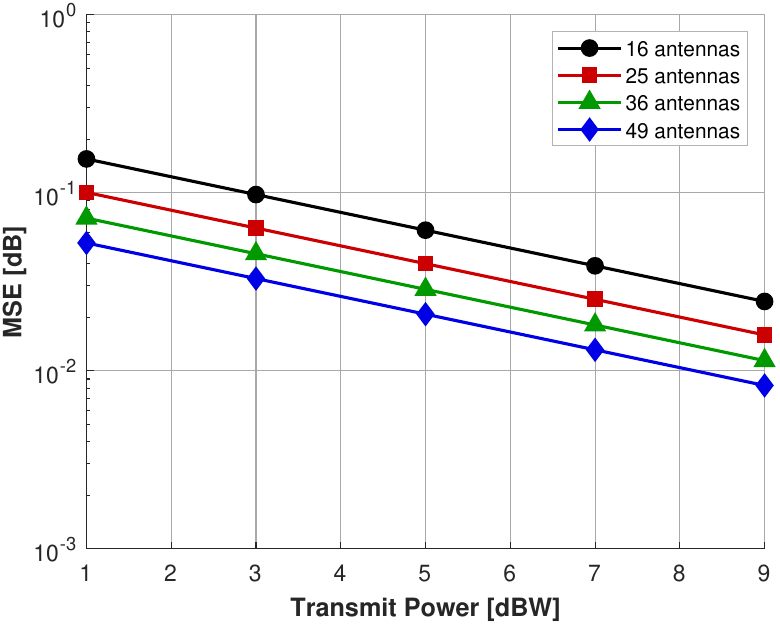}
     \caption{Bistatic radar sensing performance as a function of transmit power.}
     \label{fig5}
 \end{figure}

In Figure~\ref{fig4}, the achievable communication rate is illustrated as a function of transmit power for different numbers of antenna elements. The results show how increasing both the number of transmit antennas and the transmit power yields significant performance gains in communication throughput. Figure~\ref{fig5} presents the bistatic radar detection performance for an airplane in terms of mean-square-error (MSE), where similar improvements are observed for sensing accuracy. These results demonstrate the feasibility of the selected use case, highlighting the effectiveness of bistatic radar for joint communication and sensing. It is noteworthy that the same scenario has also been examined in the monostatic radar configuration; however, the extremely large path loss in that setting renders the use case infeasible. 

\section{Challenges and Future Research Directions} \label{sezione5}
In this section, we delve into the primary challenges that must be addressed to lay a robust foundation for the development of next-generation JCAS SN.

\subsection{Waveform Design}
Waveform design is a fundamental challenge for realizing JCAS in SN, as it dictates both communication efficiency and sensing resolution. Unlike terrestrial systems, satellites face vast propagation distances, severe path loss, long delays, and strict power budgets, all of which constrain waveform flexibility. High relative velocities between satellites, ground users, and targets introduce significant Doppler shifts and spreads, which, if unmanaged, degrade synchronization, increase error rates, and reduce detection accuracy. Wide satellite beams also illuminate heterogeneous environments—land, sea, urban, and atmospheric—resulting in complex mixed clutter, while orbital motion continuously alters sensing geometry and invalidates common stationarity assumptions. 

To address these challenges, waveform designs must be tailored to the dynamics of satellite-based JCAS. Potential directions include Orthogonal Frequency-Division Multiplexing (OFDM) variants with adaptive subcarrier spacing to cope with large Doppler shifts caused by high orbital velocities, and chirp- or Linear Frequency Modulation (LFM)-based signaling embedded with communication symbols to support long-range detection under severe path loss. Multi-carrier schemes with adaptive nulling can help mitigate heterogeneous clutter from oceans, ice, and land observed within wide satellite beams. In addition, joint optimization of pilot and data symbols is critical for simultaneous channel estimation and target tracking in rapidly time-varying satellite links. Finally, AI-assisted adaptive waveform selection could provide robustness by reconfiguring signaling parameters in real time across diverse orbital regimes (LEO, MEO, GEO) and mission profiles.


\subsection{Beamforming} 
Beamforming in JCAS-enabled SN is inherently challenging due to the stringent constraints and dynamics of the space environment. Limited satellite power budgets make it difficult to design beamformers that can simultaneously meet the requirements of both communication and sensing, often forcing trade-offs in coverage, resolution, and throughput. The highly non-stationary environment further complicates beamforming design: satellites, users, and targets are all in relative motion, demanding continuous adjustments in beam direction, shape, and power allocation to sustain reliable communication links while preserving sensing accuracy. Additional complications arise from the severe path loss over long distances and the large round-trip delays in satellite links, which limit the effectiveness of conventional closed-loop feedback and hinder real-time adaptation of beam patterns. Together, these factors render adaptive beamforming for JCAS in SN a complex and resource-intensive problem.

Future research must therefore focus on beamforming strategies tailored specifically to JCAS in the space domain. Promising directions include predictive beamforming, where orbital dynamics and ephemeris data are used to anticipate user and target positions and proactively adjust beams; hybrid analog–digital architectures that balance fine-grained sensing resolution with power efficiency; and cooperative multi-satellite beamforming, exploiting inter-satellite links to form distributed arrays for improved angular resolution and clutter suppression while improving communications quality. AI-driven beam management could further enable autonomous adaptation to evolving conditions, while multi-beam scheduling strategies may support concurrent communication and sensing tasks across wide coverage areas. Finally, embedding beamforming design with space-aware constraints, such as spectrum regulation, cross-link coordination, and limited payload resources, will be key to unlocking scalable JCAS SN.

\subsection{Doppler Effect} 
The Doppler effect is a central
phenomenon in satellite networks and presents both challenges and opportunities for JCAS. On the communication side, high orbital velocities and large relative motions between satellites, ground terminals, and targets introduce severe Doppler shifts and spreads that distort carrier frequencies, degrade synchronization, and increase error rates. These impairments are further amplified in multicarrier systems such as OFDM, where inter-carrier interference can severely reduce throughput if not properly compensated. On the sensing side, however, the same Doppler signatures carry valuable information about target motion, enabling the estimation of velocity, trajectory, and even micro-movements critical for applications such as space traffic monitoring, debris tracking, and atmospheric sensing. The inherent duality of Doppler effects, i.e., disruptive for communications yet informative for sensing, makes their joint exploitation a defining challenge for JCAS in SN.

Future research must focus on developing algorithms that simultaneously mitigate and exploit Doppler effects. Promising directions include Doppler-resilient waveform design (e.g., orthogonal time frequency space, chirp-based multicarrier signals) that naturally tolerate high mobility while preserving sensing resolution; joint Doppler estimation and compensation schemes where sensing-derived Doppler estimates are directly fed into communication receivers for improved synchronization; and predictive Doppler models based on orbital mechanics and ephemeris data to proactively adapt carrier frequencies and beam steering. AI- and ML-driven approaches can further enhance robustness by learning Doppler dynamics under diverse orbital regimes (LEO, MEO, GEO) and environmental conditions, while cooperative multi-satellite processing may exploit distributed Doppler measurements for improved accuracy in both communication alignment and target tracking. Collectively, these research directions will be crucial for turning the Doppler effect from a disruptive impairment into a resource that strengthens both communication reliability and sensing performance in future JCAS-enabled satellite systems.

\subsection{Regulatory and privacy aspects}
 The integration of JCAS into SN raises major regulatory and privacy challenges. Current spectrum regulations separate communication and radar services, yet JCAS merges them, creating uncertainty in licensing and coordination. Wide satellite footprints further amplify the risk of interference, cross-border disputes, and power flux-density violations. In parallel, high-resolution sensing can reveal personally identifiable information (PII) or sensitive activity patterns. With global satellite coverage and frequent line-of-sight access, obtaining user consent is infeasible, highlighting the urgent need for new governance frameworks.

Future research should develop frameworks that jointly address spectrum regulation and privacy in JCAS-enabled satellite networks. On the regulatory side, function-based licensing and dynamic spectrum authorization can replace rigid service distinctions, while policy-aware waveform and beamforming controllers can enforce ITU-R and national constraints in real time. On the privacy side, privacy-by-design techniques, such as resolution throttling, spatial aggregation, on-board filtering, and federated processing, will be essential to protect sensitive information without undermining mission value. Scalable governance tools, including geofenced “no-sense” zones, purpose-limited data retention, and standardized audit trails with cryptographic provenance, are also needed. Collectively, these efforts will ensure that satellite-based JCAS develops in a way that is both technically effective and socially responsible.


\section{Conclusion} \label{sezione6}
In conclusion, this work underscores the transformative potential of JCAS in driving the evolution of satellite networks for the next-generation space era. The study begins by highlighting the motivation and promise of integrating JCAS into SN through a unified hardware platform. It then explores novel use cases enabled by this integration, demonstrating the diverse applications of JCAS in SN. As a case study, the joint communications and object detection is SN is evaluated, with results demonstrating its feasibility in the bistatic radar case. Finally, the paper discusses current challenges that must be overcome and highlight promising future research directions inherent to the space based JCAS, to fully realize its potential at the global scale.

\bibliographystyle{IEEEtran}
\bibliography{JCAS}

\begin{thebibliography}{10}
\providecommand{\url}[1]{#1}
\csname url@samestyle\endcsname
\providecommand{\newblock}{\relax}
\providecommand{\bibinfo}[2]{#2}
\providecommand{\BIBentrySTDinterwordspacing}{\spaceskip=0pt\relax}
\providecommand{\BIBentryALTinterwordstretchfactor}{4}
\providecommand{\BIBentryALTinterwordspacing}{\spaceskip=\fontdimen2\font plus
\BIBentryALTinterwordstretchfactor\fontdimen3\font minus
  \fontdimen4\font\relax}
\providecommand{\BIBforeignlanguage}[2]{{%
\expandafter\ifx\csname l@#1\endcsname\relax
\typeout{** WARNING: IEEEtran.bst: No hyphenation pattern has been}%
\typeout{** loaded for the language `#1'. Using the pattern for}%
\typeout{** the default language instead.}%
\else
\language=\csname l@#1\endcsname
\fi
#2}}
\providecommand{\BIBdecl}{\relax}
\BIBdecl

\bibitem{10396846}
W.~U. Khan, A.~Mahmood, C.~K. Sheemar, E.~Lagunas, S.~Chatzinotas, and
  B.~Ottersten, ``Reconfigurable intelligent surfaces for {6G} non-terrestrial
  networks: Assisting connectivity from the sky,'' \emph{IEEE Internet of
  Things Mag.}, vol.~7, no.~1, pp. 34--39, 2024.

\bibitem{9210567}
O.~Kodheli \emph{et~al.}, ``Satellite communications in the new space era: A
  survey and future challenges,'' \emph{IEEE Commun. Surveys \& Tuts.},
  vol.~23, no.~1, pp. 70--109, 2021.

\bibitem{giordani2020non}
M.~Giordani and M.~Zorzi, ``Non-terrestrial networks in the {6G} era:
  Challenges and opportunities,'' \emph{IEEE Network}, vol.~35, no.~2, pp.
  244--251, 2020.

\bibitem{nanoavionics2024many}
K.~NanoAvionics, ``How many satellites are in space,'' \emph{Kongsberg
  NanoAvionics. Saatavissa: https://nanoavionics.
  com/blog/how-many-satellites-are-inspace/[Viitattu: 26.01. 2024]}, 2024.

\bibitem{chepuri2023integrated}
S.~P. Chepuri, N.~Shlezinger, F.~Liu, G.~C. Alexandropoulos, S.~Buzzi, and
  Y.~C. Eldar, ``Integrated sensing and communications with reconfigurable
  intelligent surfaces: From signal modeling to processing,'' \emph{IEEE Signal
  Process. Mag.}, vol.~40, no.~6, pp. 41--62, 2023.

\bibitem{sheemar2023full}
C.~K. Sheemar, G.~C. Alexandropoulos, D.~Slock, J.~Querol, and S.~Chatzinotas,
  ``Full-duplex-enabled joint communications and sensing with reconfigurable
  intelligent surfaces,'' in \emph{Proc. EUSIPCO}, Helsinki, Finland, 2023, pp.
  1509--1513.

\bibitem{zhang2021overview}
J.~A. Zhang, F.~Liu, C.~Masouros, R.~W. Heath, Z.~Feng, L.~Zheng, and
  A.~Petropulu, ``An overview of signal processing techniques for joint
  communication and radar sensing,'' \emph{IEEE J. Sel. Topics Signal
  Process.}, vol.~15, no.~6, pp. 1295--1315, 2021.

\bibitem{sheemar2023full_magazine}
C.~K. Sheemar, S.~Solanki, E.~Lagunas, J.~Querol, S.~Chatzinotas, and
  B.~Ottersten, ``Full duplex joint communications and sensing for {6G}:
  Opportunities and challenges,'' \emph{arXiv preprint arXiv:2308.07266}, 2023.

\bibitem{sheemar2025holographic}
C.~K. Sheemar, W.~U. Khan, G.~Alexandropoulos, J.~Querol, and S.~Chatzinotas,
  ``Holographic joint communications and sensing with cramer-rao bounds,''
  \emph{Accepted in IEEE Journal on Selected Areas in Communications,
  [Available Online] arXiv preprint arXiv:2502.15248}, 2025.

\bibitem{9737357}
F.~Liu, Y.~Cui, C.~Masouros, J.~Xu, T.~X. Han, Y.~C. Eldar, and S.~Buzzi,
  ``Integrated sensing and communications: Toward dual-functional wireless
  networks for {6G} and beyond,'' \emph{IEEE J. Sel. Areas Commun.}, vol.~40,
  no.~6, pp. 1728--1767, 2022.

\bibitem{leyva2024integrated}
I.~Leyva-Mayorga, F.~Saggese, L.~Li, and P.~Popovski, ``Integrated sensing and
  communications for resource allocation in non-terrestrial networks,''
  \emph{arXiv preprint arXiv:2407.06705}, 2024.

\bibitem{naeem2022novel}
A.~Naeem, S.~Rafique, and H.~Arslan, ``A novel frame design for non-terrestrial
  network based integrated sensing and communication,'' in \emph{Proc. IEEE
  PIMRC}, 2022, pp. 577--582.

\bibitem{yin2024integrated}
L.~Yin, Z.~Liu, M.~R.~B. Shankar, M.~Alaee-Kerahroodi, and B.~Clerckx,
  ``Integrated sensing and communications enabled low earth orbit satellite
  systems,'' \emph{IEEE Network}, vol.~38, no.~6, pp. 252--258, 2024.

\bibitem{you2022beam}
L.~You, X.~Qiang, C.~G. Tsinos, F.~Liu, W.~Wang, X.~Gao, and B.~Ottersten,
  ``Beam squint-aware integrated sensing and communications for hybrid massive
  {MIMO LEO} satellite systems,'' \emph{IEEE J. Sel. Areas Commun.}, vol.~40,
  no.~10, pp. 2994--3009, 2022.

\bibitem{maral2020satellite}
G.~Maral, M.~Bousquet, and Z.~Sun, \emph{Satellite communications systems:
  systems, techniques and technology}.\hskip 1em plus 0.5em minus 0.4em\relax
  John Wiley \& Sons, 2020.

\end{thebibliography}

\end{document}